\documentclass[superscriptaddress,twocolumn,showpacs,preprintnumbers,amsmath,amssymb,nofootinbib,eqsecnum]{revtex4-1}
\usepackage{graphicx}
\usepackage{color}

\begin{document}

\title{Killing--Yano forms and Killing tensors on a warped space}

\author{Pavel Krtou\v{s}}

\email{Pavel.Krtous@utf.mff.cuni.cz}

\affiliation{Institute of Theoretical Physics,
Faculty of Mathematics and Physics, Charles University,
V~Hole\v{s}ovi\v{c}k\'ach~2, Prague, 18000, Czech Republic}

\author{David Kubiz\v n\'ak}

\email{dkubiznak@perimeterinstitute.ca}

\affiliation{Perimeter Institute, 31 Caroline Street North, Waterloo, ON, N2L 2Y5, Canada}


\author{Ivan Kol\'{a}\v{r}}

\email{Ivan.Kolar@utf.mff.cuni.cz}

\affiliation{Institute of Theoretical Physics,
Faculty of Mathematics and Physics, Charles University,
V~Hole\v{s}ovi\v{c}k\'ach~2, Prague, 18000, Czech Republic}

\date{August 11, 2015} 

\begin{abstract}
We formulate several criteria under which the symmetries associated with the Killing and Killing--Yano tensors on the base space can be lifted to the symmetries of the full warped geometry. The procedure is explicitly illustrated on several examples, providing new prototypes of spacetimes
admitting such tensors. In particular, we study a warped product of two Kerr-NUT-(A)dS spacetimes and show that it gives rise to a new class of highly symmetric vacuum (with cosmological constant) black hole solutions that inherit many of the properties of the Kerr-NUT-(A)dS geometry.

\end{abstract}


\maketitle

\section{Introduction}
Introduced into physics by Penrose et. al. in the early seventies \cite{Walker:1970un, Hughston:1972qf, penrose1973naked},
the totally symmetric {\em Killing tensors} and the skew symmetric {\em Killing--Yano} forms have played an important role for the developments of gravitational and mathematical physics ever since then.
For example, they find a wide variety of applications in classical and quantum physics \cite{Cariglia:2014ysa}, play an important role for studying various integrability properties of (higher-dimensional) black holes \cite{Frolov:2008jr, Yasui:2011pr}, are related to special manifolds \cite{Semmelmann:2002fra}, or naturally appear in the context of string theory \cite{Chow:2008fe, Chervonyi:2015ima}.
Since, contrary to  Killing vectors, Killing and Killing--Yano tensors do not have clear geometrical meaning---they no longer describe continuous symmetries of the geometry but can rather be considered as `symmetries of the phase space'---they are sometimes called dynamical, or {\em hidden symmetries}.


The existence of hidden symmetries imposes strict restrictions on the background geometry, e.g. \cite{collinson1974existence, dietz1981space, Mason:2010zzc, Krtous:2008tb, Houri:2008th, Houri:2014hma}. Consequently, not every manifold admits such symmetries. Even if the symmetries are present, finding their form {\em explicitly} by solving the corresponding differential equations is a formidable task.  For this reason, it is of extreme value to seek alternative ways for finding such symmetries.

In this paper we proceed in this direction. Namely, we study hidden symmetries on {\em warped spaces}, formulating various criteria under which the Killing--Yano and Killing tensors on the base space can be {\em lifted} to symmetries of the full warped geometry. (For a different type of lift, see e.g. \cite{Cariglia:2012fi}.)
This decomposes a task of finding such symmetries
to a smaller problem (that of finding hidden symmetries for a smaller seed metric) and opens a way towards extending the applicability of hidden symmetries to more complicated spacetimes. The procedure is illustrated on several examples of physical interest, including rotating black strings, higher-dimensional singly spinning Kerr-AdS metrics, and the 5-dimensional Eguchi--Hanson soliton. We shall also consider a warped product of two Kerr-NUT-(A)dS geometries, constructing thus a {\em new class} of highly symmetric vacuum black hole solutions admitting towers of hidden symmetries lifted from the seed metrics.

Our paper is organized as follows. In the next section we review the basic definitions of Killing and Killing--Yano tensors. In Sec.~III four Theorems for lifting the seed symmetry to the full warped geometry are formulated. A concrete application of these theorems is illustrated for several examples in Secs.~IV and V. The results are summarized in Sec. VI. The appendices contain supplementary material: Appendix~\ref{apx:proof} is devoted to the proofs of the Theorems, Appendix~\ref{apx:onshellwarp} contains additional material about the Kerr-NUT-(A)dS spaces and their warped product.

\section{Killing--Yano and Killing tensors}

\subsection{Notations}
Let us briefly explain our notations.
In what follows we try to avoid writing explicitly the tensor indices, highlighting the tensor character of objects by boldface.
At the same time, we adopt a standard convention and do not distinguish tensors with covariant and contravariant indices---indices are automatically lowered or raised using the metric $\tens{g}$ or the inverse metric $\tens{g}^{-1}$, respectively.
The dot ``${\cdot}$'' between two objects represents a (one index) contraction. For example, the divergence $\nabla\cdot \tens{h}$ is given by
${(\nabla\cdot\tens{h})_{a_1\dots a_p}=\nabla^n\tens{h}_{na_1\dots a_p}}$.

We shall also employ the symmetric and multiple contraction products.
Namely, operation``$\vee$'' stands for the normalized {\em symmetric} tensor product,
\be
({\tens{A}\vee\tens{B})^{a_1\dots a_rb_1\dots b_s}} = {\binom{r+s}{r} \tens{A}^{(a_1\dots a_r}\tens{B}^{b_1\dots b_s)}}\,.
\ee
The {\em multiple contraction}, ``$\pbullet{r}$'', denotes the partial contraction of two antisymmetric forms in the first $r$ indices divided by $r!$,
\begin{equation}
  (\tens{\alpha}\pbullet{r}\tens{\beta})_{a{\dots}b{\dots}}
    =\frac1{r!}\,\tens{\alpha}_{n_1\dots n_ra\dots}\,\tens{\beta}^{n_1\dots n_r}{}_{b\dots}\;.
\end{equation}
If no index $r$ is indicated, the full contraction in all indices is assumed.
In terms of this operation we can also write down the {\em Hodge dual} as
\begin{equation}\label{Hdual}
   \hodge\tens{\alpha} = \tens{\alpha}\bullet\LCT\;,
\end{equation}
where $\LCT$ stands for the Levi-Civita tensor.

\subsection{Review of definitions}

The ${p}$-form ${\tens{f}}$ is a \emph{Killing--Yano (KY)} form \cite{yano1952some} if there exists a \mbox{${(p{+}1)}$-form} ${\tens{\ph}}$ such that
\begin{equation}\label{KYeq}
    \covd_{\!\textstyle\tens{a}}\tens{f} = \tens{a}\cdot\tens{\ph}\,
\end{equation}
for any vector ${\tens{a}}$. The `{\em strength}' ${\tens{\ph}}$ of the KY form is then uniquely determined as
\begin{equation}\label{KYstr}
    \tens{\ph} = \frac1{p+1}\grad\tens{f}\;.
\end{equation}
The ${q}$-form $\tens{h}$ is a \emph{closed conformal Killing-Yano (CCKY) form} if there exists a ${(q{-}1)}$-form ${\tens{\xi}}$ so that
\begin{equation}\label{CCKYeq}
   \nabla_{\!\textstyle\tens{a}} \tens{h} = \tens{a}\wedge \tens{\xi}\;
\end{equation}
holds for an arbitrary vector $\tens{a}$. The {strength} ${\tens{\xi}}$ is then given by
\begin{equation}\label{CCKYstr}
    \tens{\xi} = \frac{1}{D-q+1} \covd\cdot \tens{h}\;,
\end{equation}
where $D$ stands for the number of spacetime dimensions.
Note that the notions of KY and CCKY forms are Hodge dual: the Hodge dual ${\tens{h}=\hodge\tens{f}}$ of a KY form ${\tens{f}}$ with the strength ${\tens{\ph}}$ is a CCKY form with the strength ${\tens{\xi}=-\hodge\tens{\ph}}$, and vice versa.

The \emph{Killing tensor} ${\tens{k}}$ of rank ${r}$ is a totally symmetric tensor satisfying
\begin{equation}\label{KTeq}
    \covd\vee\tens{k} = 0\;.
\end{equation}
The totally symmetric tensor $\tens{q}$ is a \emph{conformal Killing tensor (CKT)} of rank $r$ if there exists a symmetric tensor $\tens{\sigma}$ of rank $r{-}1$ such that
\begin{equation}\label{CKTeq}
    \covd\vee\tens{q}= \mtrc\vee\tens{\sigma}\;.
\end{equation}
We say that the strength $\tens{\sigma}$ determines the symmetric derivative of CKT $\tens{q}$. It can be expressed in terms of the metric traces of $\covd\vee\tens{q}$. For example, for a rank 2 CKT it reads
\begin{equation}\label{CKTvect}
    \tens{\sigma} = \frac1{D+2}\bigl(2\covd\cdot\tens{q}+\covd q\bigr)\;,\quad q= \tens{q}^n_n\;.
\end{equation}
If $\tens{\sigma}$ is exact, $\tens{\sigma}=\grad A$, the CKT $\tens{q}$ defines a Killing tensor $\tens{k}$ according to \cite{Walker:1970un, Hughston:1972qf}
\begin{equation}\label{KT-CKT}
    \tens{k} +\tens{q}= A\, \mtrc^{-1}\;.
\end{equation}

Partially contracted squares of KY and CCKY forms generate Killing tensors and CKTs of rank 2, respectively.\footnote{Note that the converse  is not true: not every (conformal) Killing tensor can be written as a square of a (conformal) KY tensor.}
Namely, a KY form $\tens{f}$ of rank $p$ defines a second rank Killing tensor
\begin{equation}\label{fsq}
    \tens{k} = \tens{f}\pbullet{p{-}1}\tens{f}\;.
\end{equation}
Similarly, a rank $p$ CCKY form $\tens{h}$ defines a CKT
\begin{equation}\label{hsq}
    \tens{q} = \tens{h}\pbullet{p{-}1}\tens{h}\;.
\end{equation}
For the Hodge dual forms $\tens{f}=\hodge\tens{h}$ these tensors are related by \eqref{KT-CKT}, where
\begin{equation}\label{Arel}
    A=\tens{f}\bullet\tens{f}=\tens{h}\bullet\tens{h}\;.
\end{equation}

\section{Lifting the hidden symmetry}

\subsection*{Warped spaces}

The warped space $M$ can be realized as a direct product $M=\tilde{M}\times\bar{M}$ of two manifolds of arbitrary dimensions $\Dt$ and $\Db$, with the metric
\begin{equation}\label{warpedmtrc}
    \mtrc = \mtrt + \warp^2 \mtrb\;.
\end{equation}
Metrics $\mtrt$ and $\mtrb$ are called the base (seed) metrics and $\warp$ is the warp factor.
The Levi-Civita tensor has the form
\begin{equation}\label{warpedLCT}
    \LCT = \warp^\Db\; \LCTt \wedge  \LCTb\;.
\end{equation}
If not said otherwise, we assume that tilded objects ${\tilde{\tens{A}}}$ are non-trivial only in ``tilded directions'' and, similarly, barred objects ${\bar{\tens{A}}}$ only in ``barred directions''. The orthogonal splitting of the tangent tensor spaces $\mathbf{T}M=\mathbf{T}\tilde{M}\oplus\mathbf{T}\bar{M}$ is well defined thanks to the diagonal character of the metric \eqref{warpedmtrc}. Also, if not said otherwise, the tilded and barred objects  will depend only on a position in  $\tilde{M}$ or $\bar{M}$, respectively.

The curvature tensor of a warped space can be reconstructed in terms of the curvatures of the seed metrics and the so called \emph{Hessian tensor $\Hess$} \cite{dobarro2005curvature}. For the Ricci tensor and the scalar curvature one gets
\ba
\Ric &=& \Rict + \Ricb -\frac\Db\warp \Hess - \bigl[\frac{1}{\warp}\hess+(\Db-1)\lFsq\bigr]\warp^2\mtrb\;,\label{RicciWarp}\nonumber\\
\scR &=& \scRt +\frac1{\warp^2}\scRb - \frac{2\Db}{\warp} \hess -\Db(\Db-1)\lFsq\;,\label{scRWarp}
\ea
where
\ba
\Hess &=& \covdt\covdt\warp\;,\quad\hess=\Hess_{mn}\mtrt^{mn}\;,\label{Hess}\nonumber\\
\lF &=& \frac1\warp\grad\warp\;,\quad
  \lFsq = \mtrt^{ab}\lF_a\lF_b\;.\label{lF}
\ea
As shown in the next subsection, the {\em logarithmic gradient} $\lF$ of the warp factor $\warp$ plays an important role for the lift of hidden symmetries.

\subsection{Lifting theorems}
In this section we formulate several ``symmetry lifting constructions'', where a symmetry of the base space is lifted to a symmetry of the full warped geometry. While results of Theorems 1 and 2 are already partially known in the literature, see e.g. \cite{Kubiznak:2009sm}, Theorems 3 and 4 are, we believe, entirely new. The proof of each Theorem can be found in 
Appendix~\ref{apx:proof}.

Let us first concentrate on the seed metric ${\mtrb}$.\\
\textbf{Theorem 1.} {\it
Let the seed  metric ${\mtrb}$ of the warped geometry \eqref{warpedmtrc} admits a KY $p$-form ${\fFb}$ or a CCKY $q$-form ${\hFb}$.
Then the following forms:
\ba
\tens{f} &=& \warp^{p{+}1}\fFb\,, \label{KYliftb}\\
\tens{h} &=& \warp^{q{+}1}\LCTt\wedge\hFb\,, \label{CCKYliftb}
\ea
are the KY $p$-form or the CCKY ${(\Dt{+}q)}$-form
of the full warped geometry \eqref{warpedmtrc}.
}
\\
\textbf{Theorem 2.} {\it
If ${\bar{\tens{k}}}$ is a rank $r$ Killing tensor of the metric ${\mtrb}$, then
\begin{equation}\label{KTliftb}
    \tens{k}^{a_1\dots a_r} = \bar{\tens{k}}^{a_1\dots a_r}
\end{equation}
is a Killing tensor of the full warped geometry ${\mtrc}$.}

A similar construction can be formulated for the symmetries of metric ${\mtrt}$. However, in this case additional conditions on the warp factor ${\warp}$ have to be satisfied.
\\
\textbf{Theorem 3.} {\it
Let ${\fFt}$ be a KY ${p}$-form of the seed metric ${\mtrt}$ of the warped geometry \eqref{warpedmtrc} and let the warped factor ${\warp}$ satisfies
\begin{equation}\label{warpKYcond}
    \tilde{\grad} \bigl(\warp^{-(p{+}1)}\fFt\bigr)=0\;.
\end{equation}
Then  the following ${(\Db+p)}$-form:
\begin{equation}\label{KYliftt}
    \tens{f}=\warp^\Db \fFt\wedge\LCTb
\end{equation}
is a KY form of the full metric \eqref{warpedmtrc}.
Similarly, let ${\hFt}$ be a CCKY \mbox{${q}$-form} of ${\mtrt}$ and the the warp factor satisfies
\begin{equation}\label{warpCCKYcond}
    \covdt\cdot \bigl(\warp^{-(\Dt{+}q{+}1)}\hFt\bigr)=0\;.
\end{equation}
Then the following ${q}$-form:
\begin{equation}\label{CCKYliftt}
    \tens{h}=\hFt
\end{equation}
is a CCKY form of the metric \eqref{warpedmtrc}.
}

The conditions \eqref{warpKYcond} and \eqref{warpCCKYcond} can be written in a different form. Using the Leibnitz rule, we find
\begin{gather}
    \tilde{\grad}\fFt = (p+1)\,\lF\wedge\fFt\;,\label{warpKYcondalt1}\\
    \covdt\cdot \hFt = (\Dt-q+1)\,\lF\cdot\hFt\;,\label{warpCCKYcondalt1}
\end{gather}
where the closed 1-form ${\lF}$ is defined in \eqref{lF}. Employing the KY and CCKY strengths \eqref{KYstr} and \eqref{CCKYstr}, respectively, one obtains
\begin{equation}\label{warpKYCCKYcondalt2}
    \tilde{\tens{\ph}} = \lF\wedge\fFt\;,\quad
    \tilde{\tens{\xi}} = \lF\cdot\hFt\;.
\end{equation}
The conditions are equivalent under the Hodge duality ${\hFt=\hodget\fFt}$ with ${q=\Dt-p}$.

\textbf{Theorem 4.} {\it
Let ${\tilde{\tens{q}}}$ be a rank 2 CKT of the metric ${\mtrt}$ with its symmetric derivative given by vector $\tilde{\tens{\sigma}}$,
\eqref{CKTvect}, and the logarithmic gradient $\lF = \warp^{-1}\tilde{\grad}\warp$ of the warp factor satisfies
\begin{equation}\label{warpCKTcond}
    \tilde{\tens{\sigma}}=2\,\tilde{\tens{q}}\cdot\lF\;.
\end{equation}
Then
\begin{equation}\label{CKTliftb}
    \tens{q}^{ab} = \tilde{\tens{q}}^{ab}
\end{equation}
is a CKT of the warped metric ${\mtrc}$ and its symmetric derivative is given by vector $\tens{\sigma}^a=\tilde{\tens{\sigma}}^a$.
}

Theorems 1 and 2 are related in a sense of \eqref{fsq}. Similarly, Theorems 3 and 4 are compatible with the relation \eqref{hsq}: if a CCKY form $\tilde{\tens{h}}$ satisfies the condition \eqref{warpCCKYcond}, its square $\tilde{\tens{q}}$, \eqref{hsq}, satisfies the condition \eqref{warpCKTcond} and the CKT $\tens{q}$ on the warped space is a square of CCKY form $\tens{h}$. However, the theorems in these pairs are not equivalent or the latter is not a corollary of the former, since the existence of a Killing tensor (or a CKT) not necessarily requires the existence of its KY (CCKY) ``square root''.

\section{Three simple examples}
As a first simple example, let us consider a {\em rotating black string} in five dimensions. The
metric takes the form \eqref{warpedmtrc}, with a trivial warp factor $\warp=1$, $\mtrt=\grad z^2$, and
$\mtrb$ given by the Kerr geometry,
\ba\label{Kerr}
\mtrb&=&
-\frac{\Delta}{\rho^2}\left[\grad t - {a} \sin^2\theta \grad\phi \right]^2 + \frac{\rho^2}{\Delta}\grad r^2 + {\rho^2} \grad\theta^2\qquad \nonumber \\
&+& \frac{\sin^2\theta }{\rho^2} \left[a \grad t - {(r^2+a^2)} \grad \phi \right]^2\,,
\ea
where
\ba
\Delta &=& r^2+a^2 - 2mr\,,\quad \rho^2 = r^2+a^2\cos^2\theta\,.
\ea
The Kerr metric \eqref{Kerr} admits a non-trivial KY 2-form \cite{penrose1973naked}
\ba
\tens{\bar f}&=&a\cos\theta \grad r \wedge \bigl(\grad t -a\sin^2\!\theta \grad \phi\bigr)\nonumber\\
&&-r\sin\theta \grad \theta \wedge \bigl(a\grad t-(r^2+a^2)\grad \phi\bigr)\,.
\ea
Applying the Theorem 1, this immediately lifts to the KY 2-form $\tens{f}=\tens{\bar f}$ of the black string in five dimensions.

As a second non-trivial example, let us consider the
{\em singly-spinning Kerr-AdS} metric in $d$ number of dimensions \cite{Hawking:1998kw}. This metric can be written in the form
\eqref{warpedmtrc}, where, in the standard Boyer--Lindquist coordinates, we have
\ba
\mtrt&=&
-\frac{\Delta}{\rho^2}\left[\grad t - \frac{a}{\Xi} \sin^2\theta \grad\phi \right]^2 + \frac{\rho^2}{\Delta}\grad r^2 + \frac{\rho^2}{\Sigma} \grad\theta^2\qquad \nonumber \\
&+& \frac{\Sigma \sin^2\theta }{\rho^2} \left[a \grad t - \frac{r^2+a^2}{\Xi} \grad \phi \right]^2\,,\\
\mtrb&=& \grad\Omega_{d-4}^2\,,\quad \warp^2=r^2 \cos^2\theta\,,
\ea
and
\ba
\Delta &=& (r^2+a^2)(1+\frac{r^2}{l^2}) - 2mr^{5-d}, \quad \Sigma = 1-\frac{a^2}{l^2}\cos^2\theta\,,\nonumber\\
\Xi &=& 1-\frac{a^2}{l^2}, \quad \rho^2 = r^2+a^2\cos^2\theta\,.
\ea
It is known, e.g. \cite{Kubiznak:2007kh}, that $\mtrt$ admits a non-trivial CCKY 2-form, given by
$\tens{\tilde h}=\grad \tens{b}$, where
\be
2\tens{b}=(r^2+a^2\sin^2\!\theta)\grad t-\frac{a}{\Xi}\sin^2\!\theta(a^2+r^2){\grad \phi}\,.
\ee
One can easily verify the validity of condition \eqref{warpCCKYcond}. The Theorem 3 then implies that $\tens{h}=\tens{\tilde h}$ is a CCKY 2-form of the
full $d$-dimensional Kerr-AdS geometry.

Finally, consider the five-dimensional {\em Eguchi--Hanson soliton} \cite{Clarkson:2006zk}. The metric writes as \eqref{warpedmtrc}, with
\ba
\mtrt&=&
-\Delta\left[\grad t +2n\cos\theta \grad\phi \right]^2 + \frac{\grad r^2}{\Delta}+(r^2+n^2)\grad \Omega_2^2\,,\qquad\\
\mtrb&=& \grad z^2\,,\quad \warp^2=r^2\,,\quad \Delta=\frac{4ml^2-2n^2r^2-r^4}{l^2(r^2+n^2)}\,,\quad
\ea
and $n^2=l^2/4$.
The metric $\mtrt$ admits a non-trivial CCKY 2-form, given by
$\tens{\tilde h}=\grad \tens{b}$, where \cite{Kubiznak:2007kh}
\be
2\tens{b}=r^2\grad t+2n(r^2+n^2)\cos(\theta)\grad \phi\,.
\ee
Again, one can easily check the condition \eqref{warpCCKYcond}. The Theorem 3 then implies that $\tens{h}=\tens{\tilde h}$ is a CCKY 2-form of the
full 5-dimensional Eguchi--Hanson soliton.

\section{Warping Kerr-NUT-(A)dS}\label{warping}

A prominent example of a geometry with more than one hidden symmetry is the general Kerr-NUT-(A)dS spacetime \cite{Chen:2006xh}.
In an even dimension, $D=2N$,\footnote{%
We concentrate on even dimensions for simplicity of illustration. The same construction applies to odd dimensions as well, with additional ``odd terms'' present.}
the metric, written in the Carter-like coordinates ${x_\mu}$, ${\mu=1,\dots,N}$, and  ${\psi_i}$, ${i=0,\dots,N-1}$, reads
\begin{equation}\label{KerrNUTAdS}
    \mtrc = \sum_\mu\Biggl[\,\frac{\U_\mu}{X_\mu}\,\grad x_\mu^2+\frac{X_\mu}{\U_\mu}\biggl(\sum_k \A{k}_\mu\grad\psi_k\biggr)^{\!\!2}\,\Biggr]\;,
\end{equation}
where
\ba\label{AUdef}
\A{k}_\mu &=&\sum_{\substack{\nu_1,\dots,\nu_k\\\nu_1<\dots<\nu_k\\\nu_i\neq\mu}}
x_{\nu_1}^2\dots x_{\nu_k}^2\;,\quad
\U_\mu = \prod_{\substack{\nu\\\nu\neq\mu}}(x_\nu^2-x_\mu^2)\;,\nonumber\\
 \A{k} &=& \sum_{\substack{\nu_1,\dots,\nu_k\\\nu_1<\dots<\nu_k}}
        x_{\nu_1}^2\dots x_{\nu_k}^2\;.
\ea
The metric is {\em on-shell}, that is, it obeys the vacuum Einstein equations with the cosmological constant $\Lambda=(2N-1)(N-1)c_N$, provided the metric
functions $X_\mu$ take the following special polynomial form:
\begin{equation}\label{onshellX}
    X_\mu = \sum_{k=0}^{N} c_k (-x_\mu^2)^{k} - 2\NUT_\mu x_\mu\;.
\end{equation}
In particular, for a black hole solution (provided proper ranges and Wick rotations of coordinates are chosen and some relations between parameters are satisfied), parameters $c_j$, $j=0,\dots,N-1$ are related to rotations, while parameters $b_\mu$ are related to mass, NUT and twist charges, see \cite{Krtousetal:2015inprep} for a more detailed discussion.

Interestingly, irrespective of the metric signature or field equations, for any metric functions of the form
\be
X_\mu=X_\mu(x_\mu)\,,
\ee
the so called {\em off-shell} metric \eqref{KerrNUTAdS} admits a rich structure of hidden symmetries.
These symmetries can be generated from the \emph{principal CCKY tensor ${\tens{h}}$}, given by \cite{Kubiznak:2006kt}
\begin{equation}\label{PCCKY}
    \PCCKY = \frac12 \sum_{k}\grad\A{k+1}\wedge\grad\psi_k\;.
\end{equation}
In particular, we find the towers of symmetries summarized in the following table \cite{Krtous:2006qy}:
\be\label{A1}
\begin{array}{|l|c|}
\hline
\multicolumn{2}{|c|}{\mbox{Hidden symmetries of Kerr-NUT-(A)dS}} \\
\hline
\mbox{CCKY forms} & \CCKY{j} = \frac1{j!} \PCCKY^{\wedge j}\\
\hline
\mbox{KY forms} & \KY{j} = \hodge\CCKY{j}\\
\hline
\mbox{Killing tensors}  &  \ \KT{j} = \KY{j}\pbullet{2N{-}2j{-}1}\KY{j}\; \ \\
\hline
\mbox{CKTs}  & \CKT{j} = \CCKY{j}\pbullet{2j{-}1}\CCKY{j}\;\\
\hline
\end{array}
\ee
Forms and tensors in this tower are defined for $j=0,\dots,N$, however, for some values of~ $j$, they are trivial. Namely, ${\CCKY{0}=1}$, ${\CCKY{N}=\sqrt{\A{N}}\LCT}$, ${\KY{0}=\LCT}$,  ${\KY{N}=\sqrt{\A{N}}}$, ${\KT{0}=\mtrc^{-1}}$, ${\KT{N}=0}$, ${\CKT{0}=0}$, ${\CKT{N}=\A{N}\mtrc^{-1}}$. Note also that the the Killing tensors are related to the CKTs by \cite{Krtous:2006qy}
\begin{equation}\label{KTCKTrel}
    \KT{j}+\CKT{j}=\A{j}\mtrc^{-1}\;,
\end{equation}
reflecting the fact that the strengths of CKTs are given by
\be\label{sigmajj}
\tens{\sigma}_{(j)} = \mtrc^{-1}\cdot\grad\A{j}\;.
\ee

Let us now construct an example of a {\em new warped geometry} \eqref{warpedmtrc} where the two bases ${\mtrt}$ and ${\mtrb}$ are the off-shell Kerr-NUT-(A)dS geometries \eqref{KerrNUTAdS} of dimension ${\Dt=2\Nt}$ and ${\Db=2\Nb}$, respectively, while we choose the following
warp factor:
\begin{equation}\label{KerrNUTAdswarp}
    \warp^2 = \At{\Nt} = \xt_1^2\dots\xt_\Nt^2 \;.
\end{equation}
Obviously, such a warped space is not an off-shell Kerr-NUT-(A)dS geometry \eqref{KerrNUTAdS}. However, as we shall see, it shares some  important properties with the Kerr-NUT-(A)dS geometry. Namely, it possesses the tower of hidden symmetries, that can be obtained by lifting the symmetries of the two seed metrics using Theorems 1--4.

Moreover, as shown in Appendix~\ref{WarpedOnshell} the warped geometry solves the vacuum Einstein equations with the cosmological constant $\Lambda$, provided we set
\begin{align}
    \Xt_\mut &= \sum_{\kt=0}^{\Nt} \tilde{c}_\kt\, (-\xt_\mut^2)^\kt
       - \frac{2\NUTt_\mut}{\xt_\mut^{2\Nb{-}1}}\;,\label{onshellXt}\\
    \Xb_\mub &= \sum_{\kb=0}^{\Nb} \bar{c}_\kb\, (-\xb_\mub^2)^\kb
       - 2\NUTb_\mub \xb_\mub\;,\label{onshellXb}
\end{align}
with
\begin{equation}\label{onshellcoef}
    \tilde{c}_\Nt= \frac{\Lambda}{(2\Nt-1)(\Nt-1)}\;,\quad
    \tilde{c}_0 = \bar{c}_\Nb 
    \;.
\end{equation}
We note that the barred metric itself is an on-shell Kerr-NUT-(A)dS geometry. However, the tilded metric has modified metric functions, with the exponent of $\xt_\mu$ in \mbox{$\NUTt_\mut$-term} depending on the dimension of the barred part.
In particular, we observe that for a two-dimensional Lorentzian metric $\mtrt$ and vanishing parameters $\NUTb_\mub=0$ the warped space reduces to the spherical Schwarzschild--Tangherlini black hole in the dimension $D=2+\Db$.

Let us now turn to the lift of the hidden symmetries. As always, the results are valid for the off-shell warped metric, for arbitrary
$\Xt_\mut=\Xt_\mut(\tilde x_\mut)$ and $\Xb_\mub=\Xb_\mub(\bar x_\mub)$.

We start from the metric ${\mtrt}$. To implement the lifting construction given by Theorems 3 and 4, we need to demonstrate that the warp factor \eqref{KerrNUTAdswarp} and the forms ${\KYt{\kt}}$, ${\CCKYt{\kt}}$, and the CKTs $\CKTt{\jt}$ satisfy the conditions \eqref{warpKYcond}, \eqref{warpCCKYcond}, and \eqref{warpCKTcond}, respectively. We do so in Appendix~\ref{ConstCond}. The theorems then imply that the warp geometry \eqref{warpedmtrc} inherits the following KY forms, CCKY forms, and CKTs:
\begin{align}
    \KY{\jt}   &= \warp^{2\Nt} \KYt{\jt}\wedge\LCT\;,\label{KY-KYt}\\
    \CCKY{\jt} &= \CCKYt{\jt}\;,\label{CCKY-CCKYt}\\
    \CKT{\jt}  &= \CKTt{\jt}\;,\label{CKT-KYt}\quad \tilde j=0,\dots, \tilde N.
\end{align}
Note that CKTs \eqref{CKT-KYt} could be obtained as a square of \eqref{CCKY-CCKYt}, according to \eqref{hsq}. Similarly, taking square \eqref{fsq} of \eqref{KY-KYt} we can define Killing tensors $\KT{\jt}$. With the help of \eqref{warpedmtrc} and \eqref{KTCKTrel} they read
\begin{equation}\label{KT-KTt}
    \KT{\jt}=\KTt{\jt}+\At{\jt}\warp^{-2}\mtrb^{-1}\;.
\end{equation}
The Killing tensors \eqref{KT-KTt} and CKTs \eqref{CKT-KYt} satisfy \eqref{KTCKTrel} with
\begin{equation}\label{Adef2}
    \A{\jt} = \KY{\jt}\bullet\KY{\jt} = \CCKY{\jt}\bullet\CCKY{\jt}\;,
\end{equation}
which, in this case, reads
\begin{equation}\label{A-At}
    \A{\jt} = \At{\jt}\;.
\end{equation}

Turning next to the metric ${\mtrb}$, the Theorem~1 immediately implies that the warped space \eqref{warpedmtrc}
inherits the following rank ${2(\Nb-\jb)}$ KY forms ${\KY{\Nt+\jb}}$ and rank $2(\Nt+\jb)$
CCKY forms $\CCKYb{\jb}$:
\ba
\KY{\Nt+\jb} &=& \warp^{2(\Nb{-}\jb){+}1}\,\KYb{\jb}\;, \label{KY-KYb}\\
\CCKY{\Nt+\jb} &=& \warp^{2\jb{+}1}\,\LCTt\wedge\CCKYb{\jb}\;, \label{CCKY-CCKYb}
\ea
while the Theorem~2 guarantees the following Killing tensors:
\begin{equation}\label{KT-KTb}
    \KT{\Nt+\jb}=\KTb{\jb}\;,
\end{equation}
that could be also obtained by taking the square \eqref{fsq} of $\KY{\Nt{+}\jb}$. Similarly, by squaring $\CCKY{\Nt{+}\jb}$, \eqref{CCKY-CCKYb}, according to \eqref{hsq}, we can construct the following CKTs:
\begin{equation}\label{CKT-CKTb}
    \CKT{\Nt+\jb}=\warp^2\Ab{\jb}\mtrt+\CKTb{\jb}\;.
\end{equation}
Killing tensors \eqref{KT-KTb} and CKTs \eqref{CKT-CKTb} are related by \eqref{KTCKTrel} with $\A{\Nt{+}\jb}$ defined by \eqref{Arel},
\begin{equation}\label{Adef1}
    \A{\Nt{+}\jb} = \KY{\Nt{+}\jb}\bullet\KY{\Nt{+}\jb} = \CCKY{\Nt{+}\jb}\bullet\CCKY{\Nt{+}\jb}\;.
\end{equation}
However, since the warped space is not the Kerr-NUT-(A)dS geometry, instead of \eqref{AUdef} we have,
\begin{equation}\label{A-Ab}
    \A{\Nt{+}\jb} = \warp^2\Ab{\jb}\;.
\end{equation}

This concludes the lift of all hidden symmetries of the two Kerr-NUT-(A)dS spaces. 
We thus constructed the full symmetry tower for the warped space \eqref{warpedmtrc} given by the product of two Kerr-NUT-(A)dS metrics \eqref{KerrNUTAdS} with the warped factor \eqref{KerrNUTAdswarp}, cf. the symmetry tower of Kerr-NUT-(A)dS in $D=2(\bar N+\tilde N)$ dimensions \eqref{A1}. Namely, we have obtained KY and CCKY forms $\KY{j}$ and $\CCKY{j}$, Killing tensors $\KT{j}$, and CKTs $\CKT{j}$ for $j=0,\dots,N$, $N=\Nt+\Nb$.\footnote{%
Let us point out, that the lift of tilded objects with $\jt=\Nt$ and barred objects with $\jb=0$ gives the same Killing objects on the warped space with $j=\Nt$. Therefore, $N+1$ values of index $j$ is consistent with $\Nt+1$ values of index~$\jt$ and $\Nb+1$ values of index~$\jb$.}

Since the warped space does not directly belong to the Kerr-NUT-(A)dS class, the same degree of symmetry may be surprising. The reason for this is discussed in \cite{Krtousetal:2015inprep}, where the warped spaces are obtained as a particular limit of the Kerr-NUT-(A)dS geometry.

\section{Summary}

Hidden symmetries associated with the Killing--Yano and Killing tensors play an important role in modern gravitational and mathematical physics. For example, such symmetries provide a powerful tool for studying various integrability properties of higher-dimensional black holes. However, to show their presence or, even more importantly, to find such symmetries explicitly for a given metric is not a straightforward task.

In this paper we have studied a `warping construction' that will allow one to generate hidden symmetries of complicated
spacetimes, starting from the symmetries of more simple ones. Namely, we have formulated four Theorems discussing the conditions under which a hidden symmetry of the base space can be lifted to the symmetry of the full warped geometry. 
We illustrated the procedure on several examples, providing new prototypes of spacetimes admitting hidden symmetries. In particular, we have constructed a new class of vacuum black hole solutions that admit the `same' tower of hidden symmetries as the general Kerr-NUT-(A)dS spacetime. A generalization to a wider class of metrics is under investigation \cite{KolarKrtous:2015inprep}. Further generalizations of this construction as well as its applications to finding new examples of physically interesting spacetimes with hidden symmetries are left for future studies.

\section*{Acknowledgments}
This research was supported in part by Perimeter Institute for Theoretical Physics and by the Natural Sciences and Engineering Research Council of Canada. Research at Perimeter Institute is supported by the Government of Canada through Industry Canada and by the Province of Ontario through the Ministry of Research and Innovation. P.~K.\ was supported by Grant GA\v{C}R~P203/12/0118.
I.~K.\ was supported by the Charles University Grant No. SVV-260211.
P.~K.\ appreciates the hospitality of the Perimeter Institute where this work started.

\appendix

\section{Proofs of Theorems}\label{apx:proof}

When working on a warped space with the metric
\begin{equation}\label{warpedmtrc2}
    \mtrc = \mtrt + \warp^2 \mtrb\;,
\end{equation}
it is natural to use adjusted coordinates ${\tilde y_\mut}$ and ${\bar y_\mub}$ which depend only on positions in $\tilde{M}$ and $\bar{M}$, respectively. Since the metric $\mtrt$ depends only on ${\tilde y_\mut}$ and $\mtrb$ only on ${\bar y_\mub}$, the Christoffel coefficients with respect to the adjusted coordinates split as
\begin{equation}\label{Christ}
    \Gamma{}^{c}_{\!ab} = \tilde{\Gamma}{}^{c}_{\!ab} + \bar{\Gamma}{}^{c}_{\!ab} + \Lambda{}^{c}_{\!ab}\;,
\end{equation}
where ${\tilde{\Gamma}{}^{c}_{\!ab}}$ and ${\bar{\Gamma}{}^{c}_{\!ab}}$ are Christoffel coefficients of the metrics ${\mtrt}$ and ${\mtrb}$, respectively, and
\begin{equation}\label{Lambdaterm}
    \Lambda{}^{c}_{\!ab} = \tilde{\lambda}_a \bar{\delta}{}^c_b + \bar{\delta}{}^c_a \tilde{\lambda}_b
       -\tilde{g}^{ce}\tilde{\lambda}_e \warp^2 \bar{g}_{ab}\;,
\end{equation}
where 1-form $\lF$ is defined in \eqref{lF}.

\subsection*{Theorem 1}
We want to prove that ${\tens{f}=\warp^{p{+}1}\fFb}$ is a KY form of the metric ${\mtrc}$, assuming that ${\fFb}$ is a KY form of ${\mtrb}$. Substituting \eqref{KYstr} into \eqref{KYeq}, we want to show that
\begin{equation}\label{KYliftproof0}
    (p+1)\covd_{\!\textstyle\tens{a}} \tens{f} =\tens{a}\cdot\grad \tens{f}\;.
\end{equation}
Let us start evaluating the right-hand side:
\begin{equation}\label{KYliftproof1}
\begin{split}
    &\tens{a}\cdot\grad\bigl(\warp^{p{+}1}\fFb\bigr)=\\
      &\;= \warp^{p{+}1} \Bigl(\bar{\tens{a}}\cdot\fFb
          + (p{+}1)\, \tilde{\tens{a}}\cdot\lF\,\fFb
          - (p{+}1)\, \lF\wedge(\bar{\tens{a}}\cdot\fFb)\Bigr)\\
      &\;=(p+1) \warp^{p{+}1} \bigl(\covdb_{\textstyle \bar{\tens{a}}}\fFb + \tilde{\tens{a}}\cdot\lF\,\fFb
          -\lF\wedge(\bar{\tens{a}}\cdot\fFb)\bigr)\;,
\end{split}\raisetag{3ex}
\end{equation}
where we have split the vector ${\tens{a}=\tilde{\tens{a}}+\bar{\tens{a}}}$ into tilded and barred directions
and in the second equality used the fact that $\fFb$ is a KY form of $\mtrb$.
To express the left-hand side of \eqref{KYliftproof0}, we need to evaluate the covariant derivative $\covd_{\textstyle\!\tens{a}}\fFb$ employing \eqref{Christ}. With the help of \eqref{Lambdaterm}, ${\tilde{\tens{\delta}}\cdot\fFb=0}$, and using the standard action of Christoffel coefficients on a $p$-form, the contribution from the \mbox{$\Lambda^{\,c}_{ab}$-term} gives
\begin{equation}\label{KYliftproof2}
    - p\, \tilde{\tens{a}}\cdot\lF\, \fFb - \lF\wedge(\bar{\tens{a}}\cdot\fFb)\;.
\end{equation}
The left-hand side of \eqref{KYliftproof0} thus reads
\begin{equation}\label{KYliftproof3}
\begin{split}
    &\covd_{\!\textstyle\tens{a}} \bigl(\warp^{p{+}1}\fFb\bigr)=\\
      &\;= \warp^{p{+}1} \Bigl( \covdb_{\!\textstyle\bar{\tens{a}}}\fFb
          + (p{+}1)\, \tilde{\tens{a}}\cdot\lF\,\fFb
          - p\, \tilde{\tens{a}}\cdot\lF\, \fFb - \lF\wedge(\bar{\tens{a}}\cdot\fFb)\Bigr)\\
      &\;=\warp^{p{+}1} \bigl(\covdb_{\textstyle \bar{\tens{a}}}\fFb + \tilde{\tens{a}}\cdot\lF\,\fFb
          -\lF\wedge(\bar{\tens{a}}\cdot\fFb)\bigr)\;,
\end{split}\raisetag{3ex}
\end{equation}
where again we have split  ${\tens{a}=\tilde{\tens{a}}+\bar{\tens{a}}}$, used \eqref{lF}, and the fact that $\fFb$ and $\warp$ depend only on barred and tilded directions, respectively. Comparing \eqref{KYliftproof1} with \eqref{KYliftproof3} we obtain \eqref{KYliftproof0}.

The Hodge dual of a KY form is a CCKY form and vice versa. To prove the symmetry lift \eqref{CCKYliftb} for a CCKY $q$-form $\hFb$, we just have to evaluate ${\hodge\bigl(\warp^{p{+}1}\fFb\bigr)}$, where ${\hFb = \hodgeb \fFb}$ and $p=\Db-q$. Employing \eqref{warpedLCT}, and the properties of contractions and of the wedge product, we get
\begin{equation}\label{CCKYliftbproof}
\begin{split}
   &\hodge\bigl(\warp^{p{+}1}\fFb\bigr)
      = \warp^{2\Dt{-}q{+}1}\fFb\bullet(\LCTt\wedge\LCTb)=\\
      &\quad= (-1)^{\Dt(\Dt-q)}\, \warp^{2\Dt{-}q{+}1}\warp^{-2(\Dt{-}q)}\LCTt\wedge (\fFb\bulletb\LCTb)\\
      &\quad= (-1)^{\Dt(\Dt-q)}\, \warp^{q{+}1} \LCTt\wedge (\hodgeb\fFb)\;,
   \end{split}
\end{equation}
proving that  \eqref{CCKYliftb} is a CCKY form.

\subsection*{Theorem 2}
Let us proceed to the lift of a Killing tensor  \eqref{KTliftb}. Employing \eqref{Christ}, \eqref{Lambdaterm}, ${\covdt\bar{\tens{k}}=0}$, and the symmetry of $\tens{k}$, we get
\begin{equation}\label{KTliftbproof1}
\begin{split}
    &\covd_{\!a_0} \bar{\tens{k}}^{a_1\dots a_r}
      = \covdb_{\!a_0} \bar{\tens{k}}^{a_1\dots a_r}\\
        &\quad+ r\,\lF_{a_0} \bar{\tens{k}}^{a_1\dots a_r}
        - r\, \warp^2 \mtrb_{a_0b}\, \lF_{a}\mtrt^{a(a_1}\, \bar{\tens{k}}^{a_2\dots a_r)b}\;.
\end{split}
\end{equation}
Raising index $a_0$ with the metric \eqref{warpedmtrc2} and taking the symmetrization in all indices, we prove \eqref{KTeq}:
\begin{equation}\label{KTliftbproof2}
\begin{split}
    &\covd^{(a_0} \bar{\tens{k}}^{a_1\dots a_r)}
     = \covdb^{(a_0} \bar{\tens{k}}^{a_1\dots a_r)}\\
        &\quad+ r\,\lF_a\mtrt^{a(a_0} \bar{\tens{k}}^{a_1\dots a_r)}
        - r\, \lF_{a}\mtrt^{a(a_0} \bar{\tens{k}}^{a_1\dots a_r)}
      =0\;.
\end{split}
\end{equation}

\subsection*{Theorem 3}
We want to prove \eqref{CCKYeq} for a CCKY form $\hFt$ of the metric $\mtrt$ with the warp factor satisfying \eqref{warpCCKYcond} or equivalently
\eqref{warpCCKYcondalt1}. Splitting the Christoffel coefficients according to \eqref{Christ} when acting on a $q$-form~$\hFt$, using \eqref{Lambdaterm}, and $\covdb\hFt=0$, we find
\begin{equation}\label{CCKYlifttproof1}
    \covd_{\!\textstyle\tens{a}}\hFt = \covdt_{\!\textstyle\tilde{\tens{a}}}\hFt
      + \bar{\tens{\alpha}} \wedge (\lF\cdot \hFt)\,.
\end{equation}
Here, ${\tens{\alpha}=\tilde{\tens{\alpha}}+\bar{\tens{\alpha}}}$ with ${\tilde{\tens{\alpha}}=\mtrt\cdot\tilde{\tens{a}}}$ and ${\bar{\tens{\alpha}}= \warp^2 \mtrb\cdot\bar{\tens{a}}}$, cf.~\eqref{warpedmtrc2}.
Now we substitute the tilded version of \eqref{CCKYeq} and the assumption \eqref{warpCCKYcondalt1} into the first term, to get
\begin{equation}\label{CCKYlifttproof2}
    \covd_{\!\textstyle\tens{a}}\hFt
      = \tilde{\tens{\alpha}}\wedge(\lF\cdot\hFt)
        + \bar{\tens{\alpha}} \wedge (\lF\cdot \hFt)
      = \tens{\alpha}\wedge (\lF\cdot \hFt) \,.
\end{equation}
Tearing off the vector $\tens{a}$ from \eqref{CCKYlifttproof1} and making the contraction, we obtain the divergence $\covd\cdot\hFt$.
Taking into account the assumption \eqref{warpCCKYcondalt1} and that $\lF\cdot\hFt$ is trivial in barred directions, the divergence $\covd\cdot\hFt$ reads
\begin{equation}\label{CCKYlifttproof3}
    \covd\cdot\hFt
      = \covdt\cdot\hFt + \Db\,(\lF\cdot \hFt)
      = (\Dt+\Db-q+1)\,\lF\cdot \hFt\;.
\end{equation}
Substituting it back to \eqref{CCKYlifttproof2} we arrive at \eqref{CCKYeq} with \eqref{CCKYstr} substituted and ${D=\Dt+\Db}$.

Next, we want to prove the symmetry lift \eqref{KYliftt} using the Hodge duality. Let $\fFt=\hodget\hFt$ is a KY form of the metric $\mtrt$. Then $\hodge\hFt$ must be a KY form of the metric $\mtrc$. Employing \eqref{warpedLCT} and the relation for the Hodge dual \eqref{Hdual}, we obtain
\begin{equation}\label{KYlifttproof1}
    \hodge\hFt
    =\warp^\Db\, \hFt \bullet (\LCTt\wedge\LCTb)
    =\warp^\Db\, (\hFt\bullett\LCTt)\wedge\LCTb
    =\warp^\Db\, (\hodget\hFt)\wedge\LCTb\;,
\end{equation}
which proves \eqref{KYliftt}.

\subsection*{Theorem 4}
Finally, we want to prove that a CKT $\tilde{\tens{q}}$ of the metric $\mtrt$ is also a CKT of the warped metric $\mtrc$, provided the condition \eqref{warpCKTcond} is satisfied. Splitting the covariant derivative using \eqref{Christ} and \eqref{Lambdaterm}, we get
\begin{equation}\label{CKTlifttproof1}
    \covd_{\!a}\tilde{\tens{q}}^{bc} = \covdt_{\!a}\tilde{\tens{q}}^{bc}
      +\bar{\tens{\delta}}_a^b\lF_{n}\tilde{\tens{q}}^{nc}
      +\bar{\tens{\delta}}_a^c\lF_{n}\tilde{\tens{q}}^{bn}\;.
\end{equation}
Raising index ${a}$ using $\mtrc$ and symmetrizing in all three indices, we find
\begin{equation}\label{CKTlifttproof2}
    \covd^{(a}\tilde{\tens{q}}^{bc)} = \covdt{}^{(a}\tilde{\tens{q}}^{bc)}
      +2\warp^{-2}\mtrb^{(ab}\tilde{\tens{q}}^{c)n}\lF_{n}\;.
\end{equation}
Taking into account that $\tilde{\tens{q}}$ satisfies tilded version of \eqref{CKTeq} and the consistency condition \eqref{warpCKTcond}, we obtain
\begin{equation}\label{CKTlifttproof3}
    \covd\vee\tilde{\tens{q}}
    = \mtrt^{-1}\vee\tilde{\tens{\sigma}} + \warp^{-2}\mtrb^{-1}\vee\tilde{\tens{\sigma}}
    = \mtrc^{-1}\vee\tilde{\tens{\sigma}}\;,
\end{equation}
which is \eqref{CKTeq} for the warped geometry we wanted to prove.


\section{Warping the Kerr-NUT-(A)dS geometry}
\label{apx:onshellwarp}
In this appendix we gather various technical results about the Kerr-NUT-(A)dS geometry and its warped product that are referred to in the main text.

\subsection{Useful identities}\label{Uiden}
Let us start by listing a couple of useful properties of functions $\A{j}_\mu$, $\U_\mu$, and $\A{j}$ defined in \eqref{AUdef}.
First, $\A{j}_\mu$ can be understood as a matrix and its inversion can be found explicitly:
\be\label{Ainvid}
    \sum_k\A{k}_\mu \frac{(-x_\nu^2)^{N{-}1{-}k}}{\U_\nu}=\delta^\nu_\mu\;,\ \,
    \sum_\mu\A{k}_\mu \frac{(-x_\mu^2)^{N{-}1{-}l}}{\U_\mu}=\delta^k_l\;.
\ee
We also have the following important lemma, e.g. \cite{Krtous:2007xg}:\\
\textbf{Lemma 1.} {\it
Functions $f_\mu=f_\mu(x_\mu)$ of one variable satisfy the condition
\begin{equation}\label{sumfu0}
    \sum_\mu\frac{f_\mu}{\U_\mu} = 0
\end{equation}
if and only if they are given by the same polynomial of degree $N-2$
\begin{equation}\label{fpol}
    f_\mu = \sum_{k=0}^{N-2} a_k x_\mu^{2k}\;.
\end{equation}}
If the right-hand side of equation \eqref{sumfu0} is non-trivial, the solution for functions $f_\mu$ is given by a particular solution plus homogeneous solution \eqref{fpol}. We list three examples which sums to a special righ-hand side:
\ba
\sum_\mu\frac{(-x_\mu^2)^{N-1}}{\U_\mu}&=&1\;,\label{sumxNa}\\
\sum_\mu\frac{(-x_\mu^2)^{N}}{\U_\mu}&=&-\A{1}\,, \label{sumxN}\\
\sum_\mu\frac1{x_\mu^2\U_\mu}&=&\frac1{\A{N}}\label{sumx-2}\;.
\ea
The last relation is useful when dealing with the warp factor \eqref{KerrNUTAdswarp}.

\subsection{More on Kerr-NUT-(A)dS}
Let us next provide more information about the general Kerr-NUT-(A)dS metric \eqref{KerrNUTAdS}. Introducing an orthonormal frame
\begin{equation}\label{1formfr}
\begin{gathered}
    \enf{\mu} = \biggl(\frac{\U_\mu}{X_\mu}\biggr)^{\!\!\frac12}\grad x_\mu\;,\quad
    \ehf{\mu} = \biggl(\frac{X_\mu}{\U_\mu}\biggr)^{\!\!\frac12}\sum_k\A{k}_\mu\grad \psi_k\;,
\end{gathered}
\end{equation}
together with its dual frame
\begin{equation}\label{vecfr}
\begin{gathered}
    \env{\mu} {=} \biggl(\frac{X_\mu}{\U_\mu}\biggr)^{\!\!\frac12}\cv{x_\mu}\;,\;\;
    \ehv{\mu} {=} \biggl(\frac{\U_\mu}{X_\mu}\biggr)^{\!\!\frac12}\sum_k\frac{(-x_\mu^2)^{N{-}1{-}k}}{\U_\mu} \,\cv{\psi_k}\;,
\end{gathered}
\end{equation}
the Kerr-NUT-(A)dS metric \eqref{KerrNUTAdS} and its inverse can be written as
\be\label{inversemetric}
\tens{g} = \sum_{\mu} \bigl(\enf{\mu}\enf{\mu}+\ehf{\mu}\ehf{\mu}\bigl)\;,\quad
\tens{g}^{-1}=\sum_{\mu} \bigl(\env{\mu}\env{\mu}+\ehv{\mu}\ehv{\mu}\bigl)\,.
\ee
The principal CCKY tensor $\tens{h}$, \eqref{PCCKY} now writes as
\be\label{PKY2}
\PCCKY = \sum_{\mu}x_\mu  \enf{\mu} \wedge \ehf{\mu}\;,
\ee
and obeys \eqref{CCKYeq} with the strength $\PKV$ given by
\begin{equation}\label{PKV}
    \PKV=\frac1{D-1}\covd\cdot\PCCKY=\mtrc\cdot\cv{\psi_0}=\sum_\mu \biggl(\frac{X_\mu}{\U_\mu}\biggr)^{\!\!\frac12} \ehf{\mu}\;.
\end{equation}
Explicit expressions for the KY and CCKY forms \eqref{A1} can be found in \cite{Krtous:2006qy}.
Killing tensors ${\KT{j}}$ and CKT $\CKT{j}$ \eqref{A1} read
\begin{align}
    \KT{j} &= \sum_{\mu} \A{j}_\mu  \Bigl(\env{\mu}\env{\mu}+\ehv{\mu}\ehv{\mu}\Bigl)\;,\label{KTframe}\\
    \CKT{j} &= \sum_{\mu} x_\mu^2\A{j{-}1}_\mu  \Bigl(\env{\mu}\env{\mu}+\ehv{\mu}\ehv{\mu}\Bigl)\;.\label{CKTframe}
\end{align}
The Levi-Civita tensor is given by
\begin{equation}\label{mtrcLCT}
  \LCT = \enf{1}\wedge\ehf{1}\wedge\dots\wedge\enf{N}\wedge\ehf{N}\;.
\end{equation}

It was shown in \cite{Hamamoto:2006zf} that the Ricci curvature is also diagonal in frame \eqref{1formfr},
\begin{equation}\label{Ricci}
  \Ric=-\sum_{\mu=1}^N \; r_\mu\; \bigl(\enf{\mu}\enf{\mu}+\ehf{\mu}\ehf{\mu}\bigl)\;,
\end{equation}
with eigenvalues
\begin{equation}\label{RicciEV}
\begin{split}
  r_\mu
    &= \frac12\,\frac{X_\mu''}{\U_\mu}
      +\sum_{\substack{\nu\\\nu\neq\mu}}\frac1{\U_\nu}\,
        \frac{x_\nu X_\nu' {-} x_\mu X_\mu'}{x_\nu^2{-}x_\mu^2}
      -\sum_{\substack{\nu\\\nu\neq\mu}}\frac1{\U_\nu}\,
        \frac{X_\nu {-} X_\mu}{x_\nu^2{-}x_\mu^2}
  \;.
\end{split}
\end{equation}
The scalar curvature simplifies to
\begin{equation}\label{sccurE}
  \scR = -\sum_{\nu} \; \frac{X_\nu''}{\U_\nu}\;.
\end{equation}
The primes denote the differentiation with respect to a single argument
of the metric function, ${X_\mu'=X_{\mu,\mu}}$.

\subsection{Consistency conditions}\label{ConstCond}

Let us now return back to the warped geometry \eqref{warpedmtrc} studied in Sec.~\ref{warping}, given by two Kerr-NUT-(A)dS seeds and the warp factor \eqref{KerrNUTAdswarp}. We want to proof the consistency conditions of Theorems~3 and 4, namely  \eqref{warpKYcondalt1}, \eqref{warpCCKYcondalt1}, and \eqref{warpCKTcond}, that guarantee that the forms ${\KYt{\kt}}$, ${\CCKYt{\kt}}$, and the CKTs $\CKTt{\jt}$ of the metric ${\mtrt}$ can be lifted to the symmetries of the full warped geometry.
Since all these conditions are formulated entirely in the language of tilded quantities, we can simplify our notations by skipping tildes in all expressions till the end of this subsection.

For the warp factor \eqref{KerrNUTAdswarp}, vector ${\mtrc^{-1}\cdot\tens{\lambda}}$ reads
\begin{equation}\label{lF-KNA}
    \mtrc^{-1}\cdot\tens{\lambda} = \sum_\mu \frac1{x_\mu}\frac{X_\mu}{\U_\mu}\cv{x_\mu}\;,
\end{equation}
cf.~\eqref{inversemetric}. Contracting it with the principal CCKY form \eqref{PKY2} and employing $\grad\A{k+1}=2x_\mu\A{k}_\mu\grad x_\mu$, we get
\begin{equation}\label{warpcndproof1}
    \tens{\lambda}\cdot\PCCKY = \sum_\mu \frac{X_\mu}{\U_\mu}\sum_k \A{k}_\mu\grad\psi_k = \mtrc\cdot\cv{\psi_0}=\PKV\;.
\end{equation}
Recalling that the contraction acts as a derivative with respect to the wedge product, the definition \eqref{A1} and the previous result \eqref{warpcndproof1} imply
\begin{equation}\label{warpcndproof2}
    \tens{\lambda}\cdot \CCKY{k} = \PKV\wedge\CCKY{k-1}\;.
\end{equation}
Acting with ${\covd}$ on $\tens{h}^{(j)}$, we find
\ba\label{warpcndproof3}
\covd^a \CCKY{k}_{a_1\dots a_{2k}}&=&
\covd^a \PCCKY_{a_1a_2} \wedge  \CCKY{k{-}1}_{a_3\dots a_{2k}}\nonumber\\
&=& \tens{\delta}^a_{a_1}\wedge\PKV_{a_2}\wedge \CCKY{k{-}1}_{a_3\dots a_{2k}}\,,
\ea
where the wedges are understood between lower indices only and, in the second equality, we have used the property \eqref{CCKYeq} of the principal CCKY form written as $\covd^a \PCCKY_{bc} = 2\tens{\delta}^a_{[b}\PKV_{c]}$.
The divergence ${\covd\cdot\CCKY{k}}$ thus reads
\begin{equation}\label{warpcndproof4}
    \covd\cdot\CCKY{k} = (D-2k+1)\,\PKV\wedge\CCKY{k{-}1}\;.
\end{equation}
Comparing \eqref{warpcndproof2} with \eqref{warpcndproof4}, we proved the condition \eqref{warpCCKYcondalt1} with ${q=2k}$. The condition \eqref{warpKYcondalt1} is equivalent to \eqref{warpCCKYcondalt1} through the Hodge duality.

To prove condition \eqref{warpCKTcond}, we contract $\tens{\lambda}=\sum_\mu\frac1{x_\mu}\grad x_\mu$ with \eqref{CKTframe} and with the help of \eqref{vecfr} we find
\begin{equation}\label{warpcndproof5}
    2\tens{\lambda}\cdot\CKT{j} = \sum_\mu 2x_\mu \frac{X_\mu}{\U_\mu} \A{j{-}1}_\mu \cv{x_\mu} = \mtrc^{-1}\cdot\grad\A{j}\;,
\end{equation}
cf.~\eqref{inversemetric} and again $\grad\A{j}=2x_\mu\A{j{-}1}_\mu\grad x_\mu$. Comparing with \eqref{sigmajj},
we proved condition \eqref{warpCKTcond}.

\subsection{On-shell warped metric}\label{WarpedOnshell}

Finally, we want to investigate for which metric functions $\Xt_{\mut}$ and $\Xb_{\mub}$ the warped geometry studied in Sec.~\ref{warping} satisfies the Einstein equations. Inspecting the structure of the Ricci tensor  \eqref{RicciWarp}, the Einstein equations with a cosmological constant ${\Ric=\frac{2\Lambda}{D-2}\mtrc}$ split into two independent parts
\begin{gather}
    \Rict = \frac{\Db}{\warp}\Hess + \frac{2\Lambda}{D-2}\mtrt\;,\quad\label{EEwarpt}\\
    \Ricb = \warp^2 \biggl(\frac{\hess}{\warp}+(\Db-1)\lFsq+\frac{2\Lambda}{D-2}\biggr)\mtrb\label{EEwarpb}\;.
\end{gather}
The geometry in space $\bar{M}$ should not depend on position in $\tilde{M}$, therefore the coefficient on the right-hand-side of \eqref{EEwarpb} must be a constant, say $K$,
\begin{equation}\label{Kdef}
    K = \warp^2 \biggl(\frac{\hess}{\warp}+(\Db-1)\lFsq+\frac{2\Lambda}{D-2}\biggr)\;.
\end{equation}
When both partial metrics are Kerr-NUT-(A)dS geometries, we can use known expressions for the covariant derivative \cite{Hamamoto:2006zf} and can calculate the Hessian explicitly. All involved quantities read
\begin{gather}
    \lFsq = \sum_\mut\frac{\Xt_\mut}{\xt_\mut^2\Ut_\mut}
    \;,\label{lFKerrNUTAdS}\\
    \Hess = \sum_\mut\frac{\warp}{2\xt_\mut}\biggl(\sum_\nut\frac{\Xt_\nut}{\Ut_\nut}\biggr)_{\!\!,\mut}
      \bigl(\envt{\mut}\envt{\mut}+\ehvt{\mut}\ehvt{\mut}\bigr)
    \;,\label{hessKerrNUTAdS}\\
    \hess = \sum_\mut\frac{\warp \Xt_\mut'}{\xt_\mut\Ut_\mut}
    \;.\quad\label{HessKerrNUTAdS}
\end{gather}
Substituting \eqref{lFKerrNUTAdS}, \eqref{hessKerrNUTAdS}, \eqref{sumxNa}, and \eqref{sumx-2} into \eqref{Kdef}, we obtain the condition
\begin{equation}\label{Kcond}
  \sum_{\mut}\frac{1}{\Ut_\mut}\biggl[
    \frac{\Xt_\mut'}{\xt_\mut}+(2\Nb-1)\frac{\Xt_\mut}{\xt_\mut^2}
    -\frac{K}{\xt_\mut^2}+\frac{\Lambda}{N-1}\bigl(-\xt_\mut^2\bigr)^{\Nt{-}1}\biggr]=0\;.
\end{equation}
Applying Lemma~1, we get
\ba\label{Xteq1}
    \xt_\mut \Xt_\mut'+(2\Nb-1)\Xt_\mut&&\nonumber\\
    &&\!\!\!\!\!\!\!\!\!\!\!\!\!\!\!\!\!\!\!\!\!\!\!\!\!\!\!\!\!\!\!\!\!\!\!\!\!
    =K+\sum_{k=0}^{\Nt-2}\tilde{a}_k x_\mut^{2(k{+}1)}
      +\frac{\Lambda}{N-1}(-\xt_\mut^2)^\Nt\;,\quad
\ea
with arbitrary coefficients $\tilde{a}_k$. The right-hand side is just a polynomial of degree $\Nt$ and the differential operator on the left-hand side is homogeneous. Solving this simple differential equation, we find (introducing minus sign in powers of $-\xt_\mut^2$ is merely convenient choice compensated by replacing constants $\tilde{a}$'s with $\tilde{c}$'s)
\be\label{Xteq2}
\Xt_\mut=\sum_{\kt=0}^{\Nt}\tilde{c}_\kt (-x_\mut^2)^\kt +\frac{\NUTt_\mut}{\xt_\mut^{2\Nb-1}}\;,
\ee
where $\tilde{b}_\mut$ are integration constants (different for each $\mut$), $\tilde{c}_\kt$, $\kt=1,\dots,\Nt-1$ are arbitrary constants (replacing~$\tilde{a}$'s), and
\begin{equation}\label{ct0N}
    \tilde{c}_\Nt = \frac{\Lambda}{(2N-1)(N-1)}\;,\quad
    \tilde{c}_0   = \frac{K}{2\Nb-1}\;.
\end{equation}
We thus found metric functions $\Xt_\mut$ for which the proportionality factor in \eqref{EEwarpb} is constant, $K=(2\Nb-1)\tilde{c}_0$. Surprisingly, recalling \eqref{Ricci}, \eqref{RicciEV}, \eqref{HessKerrNUTAdS} and identities \eqref{Ainvid}, one can show that tilded part \eqref{EEwarpt} of the Einstein equations  is already satisfied by these metric functions.

Barred part \eqref{EEwarpb} of the Einstein equations  actually requires that $\mtrb$ is on-shell Kerr--NUT--(A)dS metric with cosmological constant $(\Nb-1)K$. Following \eqref{onshellX}, the metric functions $\Xb_\mub$ are  given by
\begin{equation}\label{Xbeq}
    \Xb_\mub = \sum_{\kb=0}^{\Nb} \bar{c}_\kb\, (-\xb_\mub^2)^\kb
     - 2\NUTb_\mub \xb_\mub\;,
\end{equation}
with arbitrary constants $\NUTb_\mub$ and $\bar{c}_\kb$, provided that
\begin{equation}\label{cbN}
    \bar{c}_\Nb = \frac{K}{2\Nb-1}
    \;.
\end{equation}
Putting \eqref{ct0N} and \eqref{cbN} together, we thus derived on-shell form of the warped geometry studied in Sec.~\ref{warping}.


\begin{thebibliography}{10}

\bibitem{Walker:1970un}
M.~Walker and R.~Penrose, {\it {On quadratic first integrals of the geodesic
  equations for type [22] spacetimes}},  {\em Commun. Math. Phys.} {\bf 18}
  (1970) 265--274.

\bibitem{Hughston:1972qf}
L.~P. Hughston, R.~Penrose, P.~Sommers, and M.~Walker, {\it {On a quadratic
  first integral for the charged particle orbits in the charged kerr
  solution}},  {\em Commun. Math. Phys.} {\bf 27} (1972) 303--308.

\bibitem{penrose1973naked}
R.~Penrose, {\it Naked singularities},  {\em Annals of the New York Academy of
  Sciences} {\bf 224} (1973), no.~1 125--134.

\bibitem{Cariglia:2014ysa}
M.~Cariglia, {\it {Hidden Symmetries of Dynamics in Classical and Quantum
  Physics}},  {\em Rev. Mod. Phys.} {\bf 86} (2014) 1283,
  [\href{http://arxiv.org/abs/1411.1262}{{\tt arXiv:1411.1262}}].

\bibitem{Frolov:2008jr}
V.~P. Frolov and D.~Kubiznak, {\it {Higher-Dimensional Black Holes: Hidden
  Symmetries and Separation of Variables}},  {\em Class. Quant. Grav.} {\bf 25}
  (2008) 154005, [\href{http://arxiv.org/abs/0802.0322}{{\tt
  arXiv:0802.0322}}].

\bibitem{Yasui:2011pr}
Y.~Yasui and T.~Houri, {\it {Hidden Symmetry and Exact Solutions in Einstein
  Gravity}},  {\em Prog. Theor. Phys. Suppl.} {\bf 189} (2011) 126--164,
  [\href{http://arxiv.org/abs/1104.0852}{{\tt arXiv:1104.0852}}].

\bibitem{Semmelmann:2002fra}
U.~Semmelmann, {\it {Conformal Killing forms on Riemannian manifolds}},
  \href{http://arxiv.org/abs/math/0206117}{{\tt math/0206117}}.

\bibitem{Chow:2008fe}
D.~D.~K. Chow, {\it {Symmetries of supergravity black holes}},  {\em Class.
  Quant. Grav.} {\bf 27} (2010) 205009,
  [\href{http://arxiv.org/abs/0811.1264}{{\tt arXiv:0811.1264}}].

\bibitem{Chervonyi:2015ima}
Y.~Chervonyi and O.~Lunin, {\it {Killing(-Yano) Tensors in String Theory}},
  \href{http://arxiv.org/abs/1505.0615}{{\tt arXiv:1505.0615}}.

\bibitem{collinson1974existence}
C.~Collinson, {\it The existence of killing tensors in empty space-times.},
  {\em Tensor (NS) vol. 28 (1974), pp. 173-176} {\bf 28} (1974) 173--176.

\bibitem{dietz1981space}
W.~Dietz and R.~Rudiger, {\it Space-times admitting killing-yano tensors. i},
  in {\em Proceedings of the Royal Society of London A: Mathematical, Physical
  and Engineering Sciences}, vol.~375, pp.~361--378, The Royal Society, 1981.

\bibitem{Mason:2010zzc}
L.~J. Mason and A.~Taghavi-Chabert, {\it {Killing-Yano tensors and
  multi-Hermitian structures}},  {\em J. Geom. Phys.} {\bf 60} (2010) 907--923.

\bibitem{Krtous:2008tb}
P.~Krtou\v{s}, V.~P. Frolov, and D.~Kubiznak, {\it {Hidden Symmetries of Higher
  Dimensional Black Holes and Uniqueness of the Kerr-NUT-(A)dS spacetime}},
  {\em Phys. Rev.} {\bf D78} (2008) 064022,
  [\href{http://arxiv.org/abs/0804.4705}{{\tt arXiv:0804.4705}}].

\bibitem{Houri:2008th}
T.~Houri, T.~Oota, and Y.~Yasui, {\it {Generalized Kerr-NUT-de Sitter metrics
  in all dimensions}},  {\em Phys. Lett.} {\bf B666} (2008) 391--394,
  [\href{http://arxiv.org/abs/0805.0838}{{\tt arXiv:0805.0838}}].

\bibitem{Houri:2014hma}
T.~Houri and Y.~Yasui, {\it {A simple test for spacetime symmetry}},  {\em
  Class. Quant. Grav.} {\bf 32} (2015), no.~5 055002,
  [\href{http://arxiv.org/abs/1410.1023}{{\tt arXiv:1410.1023}}].

\bibitem{Cariglia:2012fi}
M.~Cariglia, {\it {Hidden symmetries of Eisenhart lift metrics and the Dirac
  equation with flux}},  {\em Phys. Rev.} {\bf D86} (2012) 084050,
  [\href{http://arxiv.org/abs/1206.0022}{{\tt arXiv:1206.0022}}].

\bibitem{yano1952some}
K.~Yano, {\it Some remarks on tensor fields and curvature},  {\em Annals of
  Mathematics} (1952) 328--347.

\bibitem{dobarro2005curvature}
F.~Dobarro and B.~{\"U}nal, {\it Curvature of multiply warped products},  {\em
  Journal of Geometry and Physics} {\bf 55} (2005), no.~1 75--106.

\bibitem{Kubiznak:2009sm}
D.~Kubiznak, {\it {Black hole spacetimes with Killing-Yano symmetries}},  in
  {\em {Proceedings, 16th International Congress on Mathematical Physics
  (ICMP09)}}, 2009.
\newblock \href{http://arxiv.org/abs/0909.1589}{{\tt arXiv:0909.1589}}.

\bibitem{Hawking:1998kw}
S.~W. Hawking, C.~J. Hunter, and M.~Taylor, {\it {Rotation and the AdS / CFT
  correspondence}},  {\em Phys. Rev.} {\bf D59} (1999) 064005,
  [\href{http://arxiv.org/abs/hep-th/9811056}{{\tt hep-th/9811056}}].

\bibitem{Kubiznak:2007kh}
D.~Kubiznak and P.~Krtou\v{s}, {\it {On conformal Killing-Yano tensors for
  Plebanski-Demianski family of solutions}},  {\em Phys. Rev.} {\bf D76} (2007)
  084036, [\href{http://arxiv.org/abs/0707.0409}{{\tt arXiv:0707.0409}}].

\bibitem{Clarkson:2006zk}
R.~Clarkson and R.~B. Mann, {\it {Soliton solutions to the Einstein equations
  in five dimensions}},  {\em Phys. Rev. Lett.} {\bf 96} (2006) 051104,
  [\href{http://arxiv.org/abs/hep-th/0508109}{{\tt hep-th/0508109}}].

\bibitem{Chen:2006xh}
W.~Chen, H.~Lu, and C.~N. Pope, {\it {General Kerr-NUT-AdS metrics in all
  dimensions}},  {\em Class. Quant. Grav.} {\bf 23} (2006) 5323--5340,
  [\href{http://arxiv.org/abs/hep-th/0604125}{{\tt hep-th/0604125}}].

\bibitem{Krtousetal:2015inprep}
P.~Krtou\v{s}, D.~Kubiznak, V.~P. Frolov, and I.~Kol\'{a}\v{r}, {\it {Deformed
  and twisted black holes by unspinning Kerr-NUT-(A)dS metric}},  {\em in
  preparation} (2015).

\bibitem{Kubiznak:2006kt}
D.~Kubiznak and V.~P. Frolov, {\it {Hidden Symmetry of Higher Dimensional
  Kerr-NUT-AdS Spacetimes}},  {\em Class. Quant. Grav.} {\bf 24} (2007) F1--F6,
  [\href{http://arxiv.org/abs/gr-qc/0610144}{{\tt gr-qc/0610144}}].

\bibitem{Krtous:2006qy}
P.~Krtou\v{s}, D.~Kubiznak, D.~N. Page, and V.~P. Frolov, {\it {Killing-Yano
  Tensors, Rank-2 Killing Tensors, and Conserved Quantities in Higher
  Dimensions}},  {\em JHEP} {\bf 02} (2007) 004,
  [\href{http://arxiv.org/abs/hep-th/0612029}{{\tt hep-th/0612029}}].

\bibitem{KolarKrtous:2015inprep}
I.~Kol\'a\v{r} and P.~Krtou\v{s}, {\it {Spacetimes with separable Klein--Gordon
  equation in higher dimensions}},  {\em in preparation} (2015).

\bibitem{Krtous:2007xg}
P.~Krtou\v{s}, {\it {Electromagnetic field in higher-dimensional black-hole
  spacetimes}},  {\em Phys. Rev.} {\bf D76} (2007) 084035,
  [\href{http://arxiv.org/abs/0707.0002}{{\tt arXiv:0707.0002}}].

\bibitem{Hamamoto:2006zf}
N.~Hamamoto, T.~Houri, T.~Oota, and Y.~Yasui, {\it {Kerr-NUT-de Sitter
  curvature in all dimensions}},  {\em J. Phys.} {\bf A40} (2007) F177--F184,
  [\href{http://arxiv.org/abs/hep-th/0611285}{{\tt hep-th/0611285}}].

\end{thebibliography}
%

\providecommand{\href}[2]{#2}\begingroup\raggedright\endgroup

\end{document}